\documentclass[useAMS,usegraphicx]{mn2e}

\newcommand{\gal}{\texttt{MORGANA}}
\newcommand{\gs}{\texttt{GRASIL}}

\def\lesssim{\,\lower2truept\hbox{${<\atop\hbox{\raise4truept\hbox{$\sim$}}}$}\,}
\def\gtrsim{\,\lower2truept\hbox{${>\atop\hbox{\raise4truept\hbox{$\sim$}}}$}\,}

\title[Radiative transfer with ANN]{Fast radiative transfer of dust reprocessing in semi-analytic models with artificial
neural networks}

\author[L. Silva, F. Fontanot \& G. L. Granato]{
\parbox[t]{\textwidth}{
Laura Silva$^{1}$\thanks{Email: silva@oats.inaf.it},
Fabio Fontanot$^{2,3}$\thanks{Email: fabio.fontanot@h-its.org},
Gian Luigi Granato$^{1}$\thanks{Email: granato@oats.inaf.it}}
\vspace*{6pt} \\
  $^1$ INAF-Trieste, Via G.B. Tiepolo 11, I-34131 Trieste, Italy\\
  $^2$ HITS – Heidelberger Institut f\"ur Theoretische Studien, Schloss-Wolfsbrunnenweg 35, 69118 Heidelberg, Germany\\
  $^3$ Institut f\"ur Theoretische Physik, Philosophenweg, 16, 69120, Heidelberg, Germany
}

\begin{document}
\date{Accepted ... Received ...}

\maketitle

\begin{abstract}
A serious concern for semi-analytical galaxy formation models, aiming to simulate
multi-wavelength surveys and to thoroughly explore the model parameter
space, is the extremely time consuming numerical solution of the
radiative transfer of stellar radiation through dusty media. To overcome this problem, we have implemented an artificial
neural network algorithm in the radiative transfer code \gs, in order to significantly speed up the computation of the infrared SED.
The ANN we have implemented is of general use, in that its input neurons are
defined as those quantities effectively determining the shape of the IR SED. Therefore, the training of the ANN can be performed with any model and then applied to other models. We made a blind test to check the algorithm, by applying a net trained with a standard chemical evolution model (i.e.\ \texttt{CHE\_EVO}) to a mock catalogue extracted from the SAM \gal, and compared galaxy counts and evolution of the luminosity functions
in several near-IR to sub-mm bands, and also the spectral differences for a large subset of randomly extracted models.
The ANN is able to excellently approximate the full
computation, but with a gain in CPU time by $\sim 2$ orders of magnitude.
It is only advisable that the training covers reasonably well the range of values of the input neurons in
the application. Indeed in the sub-mm at high redshift, a tiny fraction of models with some sensible input
neurons out of the range of the trained net cause wrong answer by the ANN. These are extreme starbursting models with high optical depths, favorably selected by sub-mm observations, and difficult to predict a priori.
\end{abstract}


\section{Introduction}

The great observational advances of the last decades have demonstrated the
ubiquitous presence of dust in galaxies, and its strong impact on their
Spectral Energy Distribution (SED). Dust grains efficiently absorb and scatter
radiation at short wavelengths ($\lambda<1\mu m$). The absorbed energy is
thermally re-emitted at longer wavelengths in the infrared (IR) region
($\lambda>1\mu m$). Since the early estimates based on the IRAS satellite, it has
been clear that dust in the ISM shifts from the optical-UV to the IR region at
least $\sim 30\%$ of the total power emitted by primary sources (stars and AGNs) in
nearby galaxies, and that this fraction increases with increasing galactic
activity (e.g. Sanders \& Mirabel 1996).

More recent IR and sub-mm surveys have demonstrated that the
importance of dust reprocessing increases at high redshift and
have allowed to directly map the evolution of galaxies with cosmic time
(e.g. ISO 15 $\mu$m, Elbaz et al. 1999, 2002; Gruppioni
et al. 2002; SCUBA 850 $\mu$m, Hughes et al. 1998; Chapman et al. 2005; Mortier et al. 2005;
{\it Spitzer} from from $3.6$ to $160 \mu$m, Le
Floch et al. 2005, Perez-Gonzalez et al. 2005, Babbedge et al. 2006,
Franceschini et al. 2006). The ongoing {\it Herschel} observational campaigns, with SPIRE at 250 , 350 and 500 $\mu$m,
and PACS
at 70, 110 and 170 $\mu$m are making it possible to
carry out the first large-area surveys in the almost unexplored far-IR range (e.g.\ H-ATLAS, Eales et al. 2010, and
HERMES, Oliver et al. 2010).

Dust complicates significantly the interpretation of galactic SED, in that it
introduces a dependence on the geometry and on the intrinsic and uncertain dust
properties, and calls for the solution of numerically expensive radiative
transfer problems.

The study of galaxy formation and evolution has been receiving increasing
interest also from the theoretical point of view. The modeling of galaxy
formation and evolution in cosmological context involves many processes at very
different scales, from hundreds of Mpc to much less than a pc. The widest range
of observed galaxy properties have been analyzed using the so-called
semi-analytic models (SAM; White \& Rees 1978; Lacey \& Silk 1991; White \&
Frenk 1991), that consist in calculating the evolution of the baryon component
using simple analytical approximations, involving several adjustable
parameters, while the evolution of the dark matter is calculated directly using
Monte-Carlo techniques based on the extended Press-Schechter theory,
Lagrangian perturbation theory or gravity-only N-body simulations
(see e.g.\ Lacey \& Cole 1993; Monaco, Theuns \& Taffoni 2002; Springel et al.\ 2005).

The final step to compare the SAM predictions with observations, is the computation of the
SED for each galaxy of mock galaxy catalogues, in a wavelength range as broad as possible. Indeed,
given the strong uncertainties in the recipes adopted to describe the baryon
physics, and the large set of free parameters, substantial progresses call for
a complete multi-wavelength analysis of galaxy data. The full SED of model
galaxies, at least in principle, should be calculated by appropriately taking
into account for each galaxy its particular star formation and metallicity
history and the geometrical information provided by the SAM. The
simulated SED catalogue can then be compared to real observed galaxy surveys,
so as to check whether the predictions are or are not representative of the
real universe and to retrieve some information on the galaxy formation process.

Due to the complexity and the computational cost of taking into account dust
reprocessing, most semi-analytical models have made use of simple empirical or
semi-empirical treatments (e.g. Guiderdoni et al. 1998; Kauffmann et al. 1999;
Somerville \& Primack 1999; Hatton et al. 2003; Blaizot et al. 2004; Kang et
al. 2005; Kitzbichler \& White 2007). The only SAMs that includes a UV to
sub-mm radiative transfer computed from {\it first principle}, exploiting self
consistently all the information provided by the SAM, are \texttt{GALFORM}
(Cole et al. 2000; Granato et al. 2000 [G00]; Baugh et al 2005; Lacey et al. 2008, 2010; Swinbank et
al. 2008), \texttt{MORGANA} (MOdelling the Rise of GAlaxies aNd Active nuclei,
Monaco et al. 2007; Fontanot et al. 2007, 2009a; Lo Faro et al. 2009; Fontanot \& Monaco 2010),
and \texttt{ABC} (Anti-hierarchical Baryonic Collapse; Granato et al. 2004; Silva et al. 2005;
Lapi et al. 2006). These models have been interfaced with \texttt{GRASIL} to
make detailed comparisons and predictions in different spectral ranges.

\gs\ (Silva et al. 1998 [S98]; Silva 1999 [S99]; G00;
Bressan, Silva \& Granato 2002; Panuzzo et al. 2003; Vega et
al. 2005; Schurer et al. 2009) is a relatively realistic and flexible
multi-wavelength model, which calculates a galactic SED in a reasonably short
computing time, to be applied both to interpret observations and to make
predictions in conjunction with galaxy formation models. To do this, it
includes a sufficiently realistic bulge plus disk geometry, as well as the
radiative transfer effects of different dusty environments and the clumping of
stars and dust, but avoids exceedingly time consuming Monte Carlo calculations
and allows for some degree of geometrical (axial and equatorial) symmetry. The
model has been successfully applied in many contexts (e.g. Granato et al. 2004;
Baugh et al. 2005; Silva et al. 2005; Panuzzo et al. 2007a,b; Iglesias-Paramo
et al. 2007; Fontanot et al. 2007; Galliano et al. 2008; Vega et al.
2008; Lacey et al. 2008, 2010; Michalowski et al. 2009, 2010; Schurer et al.
2009; Santini et al. 2010).

In particular, \texttt{GRASIL} has been used quite extensively for calculating
the SEDs for the above mentioned SAMs. However, the calculation of the IR SED
by \texttt{GRASIL} is still the bottleneck of the whole process of running a
realization of a SAM, limiting the usefulness of this popular approach to
galaxy formation modeling.

To improve on this, we implemented in \texttt{GRASIL} the possibility
of computing SEDs with an Artificial Neural Network (ANN) algorithm. In Silva
et al.\ (2011; henceforth Paper I) we demonstrated that this approach can
reduce the computing time by orders of magnitude, with very limited loss of
accuracy. However, in Paper I the ANN was implemented only for the two simplest
geometrical arrangements of \texttt{GRASIL}, pure disc or spherical distribution of
stars and dust, and we tested and applied it to cases suited for these
geometries. In particular, as a practical sample application in the field of
galaxy formation models, we computed galaxy counts in the {\it Herschel}
imaging bands for the \texttt{ABC} model for spheroidal galaxies. However these
simplified geometries are insufficient for a fully fledged SAM. Therefore, in
this paper, we present the implementation of the ANN for the mixed bulge+disc
geometry. To validate the technique, we compare statistical properties of
galaxies predicted by the well known SAM \gal\ (Monaco et al.\ 2007), as
derived with full \texttt{GRASIL} computations and with the ANN method.

Almeida et al. (2010) have used the ANN algorithm to insert the
\texttt{GALFORM}+\texttt{GRASIL} model into the Millennium Simulation, to study
the properties of the population of sub-mm galaxies. That method is
substantially different from the one presented here. They identify the properties
of the \texttt{GALFORM} galaxies which determine, through an ANN, their full
\texttt{GRASIL} SEDs. The method is successful and extremely fast, since the
ANN is used to compute the entire SED, not only the IR dust emission, as in our
case. However their implementation is very specific to
\texttt{GALFORM}, whose different realizations require each a
a training of the ANN, and one for each output redshift.

The method we have implemented here is less fast but far more general, because
the input is directly linked to the galactic properties effectively determining
the portion of the SED dominated by dust emission (e.g optical depths, masses
of dust, radiation field etc.), independently of any SAM. As a result, one
single training is able to cover a large variety of applications, such as to
explore the parameter space of a SAM, or to deal with different SAMs. Indeed,
we stress that {\it the ANN employed in this paper has not been trained with
SEDs computed for the semi-analytic model {\gal}}.

In Section \ref{section:grasil} we recall the main properties of \texttt{GRASIL};
in Section \ref{section:sedann} we provide some
generalities on ANN, and in Section \ref{section:anngrasil} we describe the
implementation of ANN in \texttt{GRASIL}, the choice of the input neurons and
the definition of the trained net; in Section \ref{section:morgana} we describe the main characteristics of the SAM \gal,
and in Section \ref{section:results} we show some
applications and examples. Finally our conclusions are presented in Section
\ref{section:concl}.

%%%%%%%%%%

\section{Modelling SEDs with \texttt{GRASIL}}
\label{section:grasil}

\subsection{General description}
\label{section:gendes}

\texttt{GRASIL} (\texttt{GRA}phite $\&$ \texttt{SIL}icate) is a model meant to
describe the far-UV to radio SED of galaxies, treating with particular care the
effects of dust reprocessing on the stellar radiation. Its aim is to provide a
relatively realistic and flexible modelling of galaxy SEDs, with a moderate
computing time. Below we provide a summary of its features sufficient to
understand our implementation of the ANN. We refer to the original papers for more
details (in particular S98; S99; G00;
Panuzzo et al. 2003; Vega et al.\ 2005). A somewhat more expanded summary can
be found in Paper I.

\begin{itemize}
\item Galaxies are represented with stars and dust distributed in a bulge
    and/or a disc, adopting respectively a King and a double exponential
    profile (see e.g. Fig.\ 2.7 in S99 or Fig.\ 1 in G00  for a schematic
    representation of the geometry and components).
\item Three different dusty environments and their corresponding
    interaction with stars are considered: the star-forming molecular
    clouds (MCs) associated with newly-born stars, the diffuse medium
    (''cirrus'') associated with more evolved stars, and the dusty
    envelopes around AGB stars (intermediate age stars), their relative
    contribution to the SED depending on the star formation history.
\item The birth of stars within MCs and their gradual dispersion into the
    diffuse medium is accounted for by decreasing the fraction of energy
    stars emit within MCs with increasing age, over a typical "escape
    timescale" (Sec.\ 2.5 and Eq.\ 8 in S98 for more details). Therefore we
    account for the clumping of (young) stars and dust within a diffuse
    medium, and for a greater attenuation suffered by the youngest stars
    (age-dependent attenuation)
\item The dust model is made of graphite and silicate spherical grains with
    a continuous size distribution including grains in thermal equilibrium
    with the radiation field, very small grains fluctuating in temperature
    due to the absorption of single UV photons, and PAH molecules (optical
    properties by Draine \& Lee 1984; Laor \& Draine 1993; Li \& Draine
    2001; Draine \& Li 2007). We compute the response to the incident
    radiation field for each type of grain.
\item The RT is exactly solved for the MCs with the $\Lambda$-iteration
    algorithm (e.g.\ Granato \& Danese 1994). They are represented as
    spherically symmetric clouds. Star forming MCs typically have extremely
    high optical depths even in the IR, which means IR-produced photons are
    self-absorbed, thus requiring a full RT treatment.
\item The model galaxy is binned in appropriately small volume elements.
    The radiation field is evaluated in each of them from the knowledge of
    the distribution of stars and dust. Consequently the local dust
    emission and the attenuated radiation along each desired line of sight
    is computed. The treatment of the RT and dust emission in the diffuse
    phase (the major bottleneck of the whole computation) is approximated,
    but adequate for most situations (see Section 2.5.2 in S98 and 2.5.3 in
    S99).
\item Our reference library of SSPs is from Bressan et al. (1998, 2002). We
    recall that the effects of the dusty envelopes around AGB stars and the
    radio emission (both thermal and non-thermal) are directly included in
    these SSPs. But any desired SSP library can be given in input to
    \texttt{GRASIL} (e.g. Fontanot \& Monaco 2010 tested the effects of
    both Bressan et al.\ and Maraston (2005) SSPs in
    \texttt{MORGANA}+\texttt{GRASIL}).
\end{itemize}

\subsection{Inputs}
\label{section:inpgra}

The inputs required by \texttt{GRASIL} consist of the star formation, gas and
metallicity evolution histories, and a set of geometrical parameters.
The former ingredients can be provided by analytical star formation laws, or by
``classical'' chemical evolution models, or by more complex galaxy formation
models (e.g. \texttt{GALFORM}, \texttt{MORGANA}, \texttt{ABC}, see the
Introduction).

Here, as in Paper I, we have generated the star formation histories used to
train the ANN with the code \texttt{CHE$\_$EVO} (described in S99, see also Section 2.2 in Fontanot et
al.\ 2009b). This implies that the training libraries are in principle different from
the typical \gal\ SF histories.
\texttt{CHE$\_$EVO} computes the evolution of the SFR, mass of gas,
metallicity, and of the chemical elements given an IMF and a SF law of the kind
$SFR(t)= \nu \, M_{gas}(t)^k + f(t)$, where the first term is a Schmidt-type SF
with efficiency $\nu$, and $f(t)$ is an analytical term. Note however that our
approach is independent of this choice, and indeed our aim is to compute ANNs
that can work with any engine to generate the SF histories.

The other inputs required by \texttt{GRASIL}, some of which may be provided by
galaxy formation models, are:

\begin{itemize}
\item $f_{MC}$: gas mass fraction in star forming MCs. It affects mainly
    the FIR-submm.
\item $\tau_{MC}$: optical depth of MCs. It affects strongly the mid-IR and
    the silicate absorption feature.
\item t$_{esc}$: escape time scale of young stars from the MCs. It affects
    mainly the IR to UV-optical ratio.
\item $\delta \equiv$ M$_{dust}/$M$_{gas}$: dust to gas mass ratio. It is
    customary to set it either to a fixed value or proportional to the
    metallicity, unless provided by a dust evolution model (e.g. as in
    Schurer et al. 2009).
\item Bulge scale-radii (core radii of the King profile) and disk
    scale-radii and heights (of the double exponential) for stars and dust.
    The distribution of the radiation field relative to the dust determines
    the dust temperature distribution.
\end{itemize}

%%%%%%%%%%%%%%%%%%%%%%%%%%%%%%%%%%%

\section{Computing SEDs with Artificial Neural Networks}
\label{section:sedann}

We refer the reader to Section 3 of Paper I for more details on the topic
discussed in this section. In particular, their Section 3.1 provides a summary
of basic concepts on ANNs most relevant for the present application.

In short, ANNs are computing algorithms resembling to some extent the behavior
of brains. They consist of processing units ({\it neurons} in ANN language)
with multiple connections to transmit a signal, organized as a network. These
connections have adaptable strengths ({\it synaptic weights}) which modify the
signal transmitted to and from each neuron. The {\it training} of the network
is the process of adjusting the weights, so that the network {\it learns} to
solve the specific problem at hand.

In our specific case, we have demonstrated in Paper I that the network can be trained
on a large set of pre-computed SEDs to predict, with good accuracy, the
response of the SED at each wavelength ({\it output neurons})
to variations of the physical parameters ({\it input neurons}).

\subsection{Implementing ANN in \texttt{GRASIL}}
\label{section:anngrasil}

Our strategy is to use an ANN to predict only the IR emission, both from
molecular clouds as well as cirrus, which are the true bottlenecks of \gs\
computations. The direct calculation of the extinction of starlight by the
molecular clouds and by the cirrus, which is much less demanding, provides the
dust attenuated stellar radiation. Also, it yields the amount of energy
absorbed by dust, and therefore the normalization for the two dust emission
components, while the ANN predicts their spectral shapes. Since the CPU time
required by the ANN is negligible with respect to the direct computation, we
obtain orders of magnitude gains in the time performance.

Owing to the very different nature and treatment of the MCs and cirrus
components, the quantities (input neurons) that determine their respective IR
SEDs are different. Therefore we have implemented a distinct ANN for each of
them, namely a standard feed-forward back-propagation Multi-Layer Perceptron
(MLP) with one hidden layer, using a sigmoid activation function from the input
to the hidden layer (see Paper I for definitions). To create and use
the trained net we have adapted and included in \gs\ the F90 code by B. Fiedler
freely available at {\it http://mensch.org/neural/}.

\subsection{Input and output neurons}
\label{section:neurons}

The identification of the input neurons is based on physical expectations
corroborated by working experience with \texttt{GRASIL}. As such, they are
closely related to, but not coincident with, \texttt{GRASIL} parameters. This
is because different combinations of two or more parameters produce identical
or very similar dust emission SEDs.

For instance the cirrus, and even more the MC dust emission, depends
weakly on details of the spectral shape of the input stellar radiation, which
means that different combinations of SFR$(t)$, $Z(t)$, galactic age $T_{gal}$,
and MC escape timescale $t_{esc}$ (which affects the fraction of starlight
heating the MC and that heating the cirrus) may give rise to almost
identical dust emission in one or both components.

The output neurons are the values of $\lambda L_\lambda$ in the IR region, both
for molecular clouds and for the cirrus, which usually means several hundreds
output neurons.

\subsubsection*{Input neurons for molecular clouds}

The input neurons for MCs are the same as in Paper I, since the different geometry
considered here does not affect the MC emission:

\begin{itemize}
\item $\tau_{MC} \propto \delta \,M_{MC}/R_{MC}^2$, the  molecular cloud
    optical depth (conventionally given at 1 $\mu$m).
\item $R_{MC}[pc]/(k \sqrt(L_{\star MC,46})$ the ratio of the molecular
    cloud radius over an estimate of the dust sublimation radius, i.e.\ the
    inner radius of the dust distribution. The latter depends on the
    luminosity of stars within each cloud (in $10^{46}$ erg/s). The
    constant $k$ depends on the adopted maximum temperature $T_s$ for dust,
    but its value is irrelevant as long as we use the same value when
    training and using the MLP.
\end{itemize}

\subsubsection*{Input neurons for Cirrus}

In Paper I, to cope with the cirrus emission we needed a set of 6 input neurons
for spherical symmetry and 9 for discs. For the mixed bulge and disc geometry
considered here, 3 more neurons than the latter set are necessary. In the following list we used
boldface to highlight points where there are differences or additions with
respect to Paper I:

\begin{itemize}

\item $\log (L_{Cir}/L_{\star,c})$, the cirrus dust luminosity normalized
    to the stellar luminosity heating the cirrus. The former is given by
    the amount of stellar plus MCs energy absorbed by the cirrus, from
    conservation of energy. This ratio provides a global measure of the
    amount of dust reprocessing.

\item $\log (M_{Cir}/L_{\star,c})$, the normalized cirrus dust mass,
    correlated with the (distribution of) dust emitting temperature.

\item $\tau_p$ and $\tau_e$, the polar and equatorial optical depths due to
    cirrus alone (integral of the dust density distribution along the polar
    and equatorial directions respectively, conventionally given at $1
    \mu$m).

\item $\tau_h$, a fictitious optical depth, computed as if the cirrus dust were
    spherically and homogeneously distributed. This dummy quantity measures
    the ``concentration'' of the dust distribution independently of the
    specific density law assumed. This concentration significantly affects
    the shape of the emitted SED, and we empirically found that its
    inclusion improves the performances of the MLP.

\item \textbf{Geometrical ratios}: To the three ratios used in Paper I for
    the disc geometry, $r_{d,*}/r_{d,diff}$, $z_{d,*}/r_{d,*}$,
    $z_{d,diff}/r_{d,diff}$, we added the ratio $r_{d,*}/r_{c,*}$ between
    the scalelengths of the star distributions in the disk and bulge
    component. Taken together these four ratios characterize the relative
    position of dust and stars and the geometrical thickness of star and
    dust distributions.

\item Hardness ratio: ratio of the radiation field at 0.3 $\mu$m over 1
    $\mu$m, heating the cirrus (thus emerging from molecular clouds and
    stars already out of molecular clouds). Since small grains and
    especially PAHs are excited most effectively by relatively hard UV
    photons, this quantity is correlated with the ratio between the NIR-MIR
    emission they produce, and the far-IR due to big grains.

\item \textbf{Bulge over total ratio $b/t$}: fraction of the stellar mass
    in the bulge component. This provides some information of the relative
    importance of the two stellar components in heating dust grains.

\item \textbf{Fraction of cirrus heating due to stars} and not
    to MCs: $\log (1-(L_{MC}-L_{MC,ext})/L_{Cir})$, where $L_{MC}$ and
    $L_{MC,ext}$ are the emissions of MCs before and after absorbtion by
    the diffuse cirrus dust. This was not used in Paper I, though its
    usefulness does not depend on the particular geometry. The reason is
    that, in most real situations, the cirrus emission, peaking at $\sim
    100 \mu$m, is by far not self-absorbed. Consistently, \gs\ does not deal
    with the radiative transfer of IR photons through this component. In this case even the MC
    emission, though peaking at shorter IR wavelength, where the optical
    depth is higher, is still negligibly absorbed by the cirrus. In this
    case all cirrus heating comes from starlight. However SAM can produce
    mock galaxies with thicker cirrus. In those cases in which the absorbtion of MC emission by
    cirrus is not negligible, but the cirrus self-absorbtion is still of
    little importance, \gs\ can still produce reliable SEDs. To cope with
    this with ANNs, we found empirically that it is useful to have an
    information on the fraction of cirrus heating coming from stars and MCs.
    This is likely due to the very different capability of typical
    starlight and MC photons to penetrate the diffuse cirrus distribution.
    In summary, the inclusion of this neuron improves the capability of \gs
    -ANN to work with extreme galaxies produced by SAMs.

\end{itemize}

\subsection{Network training}
\label{section:train}

The MLPs we use in the following have been trained on about $10^4$
\texttt{CHE$\_$EVO}+\texttt{GRASIL} models, likely covering the range of
parameter values arising in typical realizations of \gal:

\begin{itemize}
\item $\tau_{MC} = 0.001\ \mbox{to}\ 50$;
\item $R_{MC}/R_{min}$ = $160$ to $20000$;
\item $\log (L_{Cir}/L_{\star,c}) = -5 \ \mbox{to}\ 0.1$;
\item $\log (M_{Cir}/L_{\star,c}) = -9.84 \ \mbox{to}\  -5.84$;
\item $\tau_e = 1.6\mbox{e-5} \ \mbox{to}\  1500$;
\item $\tau_p = 2.2e\mbox{-6} \ \mbox{to}\  95$;
\item $\tau_h = 1.7e\mbox{-7} \ \mbox{to}\  1$;
\item $r_{d,*}/r_{d,diff} = 0.5 \ \mbox{to}\ 1$;
\item $z_{d,*}/r_{d,*} = 0.05 \ \mbox{to}\ 0.30$;
\item $z_{d,diff}/r_{d,diff} = 0.02 \ \mbox{to}\ 0.3$;
\item $r_{d,*}/r_{c,*} = 0.01 \ \mbox{to}\ 100$;
\item $\log (L_{\star,c}(0.3)/L_{\star,c}(1)) = -0.46 \ \mbox{to}\ 0.66$;
\item $b/t = 0.02 \ \mbox{to}\ 1$;
\item$\log (1-(L_{MC}-L_{MC,ext})/L_{Cir}) = -0.34 \ \mbox{to}\ 0$.
\end{itemize}

We trained the net with 90\% of the models, randomly chosen within the library
generated with the aforementioned range of parameters, and using the remaining
10\% as a verification set.  We have empirically adjusted the number of neurons
in the hidden layer $n_{hid}$ in the same way as in Paper I. We have adopted
$n_{hid}=20$ for molecular clouds and 47 for cirrus. We trained the MLP using
750 training epochs (iterations) with a learning rate of 0.001. We verified
that none of these choices is very critical for the final results.

An important advantage of the ANN technique with respect to classical
interpolations is the capability to ``learn'' the effect of each single input
neuron on the SED, mimicking in some sense the skill that a real
\texttt{GRASIL} user develops with experience. Therefore the MLP can
approximate sufficiently well for most purposes the SED  produced by a full
\gs\ computation, corresponding to a given choice of input neurons, even when
the trained set does not include examples with the {\it entire} set of inputs
neurons bracketing the required ones at the same time. It is often sufficient
that each single input neuron is independently within the values included for
the training.

\subsection{Computing performance}

\begin{figure}
  \centerline{
   \includegraphics[width=0.5\textwidth]{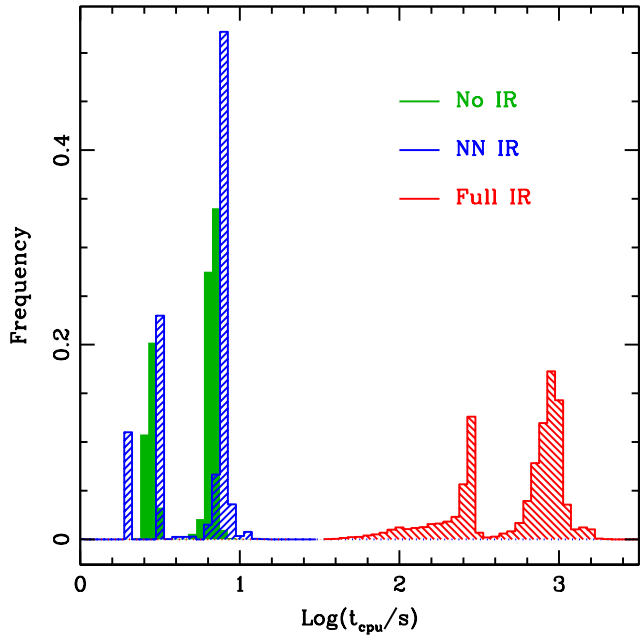}
    }
\caption{Comparison of the CPU computing time for the full- and the ANN-\texttt{GRASIL}
in the case of a typical \gal\ realization. Backslash (red) histogram is for the full computation, the forward slash (blue) one is for the ANN, and the full (green) histogram is if only extinction is computed.}
\label{fig:cputime}
\end{figure}

The implementation of the ANN into \texttt{GRASIL} dramatically reduces the CPU
time required to run the code. In Fig.\ \ref{fig:cputime} we show the histogram of the CPU
times required to compute the SED for our reference mock galaxy catalogue (see Section \ref{section:results} for more details on
the catalogue). The histograms show the computing time distributions for
a run with only optical extinction (no IR computation, solid green histograms),
a run with the full \texttt{GRASIL} computation (red backslash histograms), and
with the ANN technique (blue forward slash histogram).
The CPU gain between the full and the ANN computation is of about 2 orders of
magnitude.

%%%%%%%%%%%%%%%%%%

\begin{figure}
  \centerline{
   \includegraphics[width=0.5\textwidth]{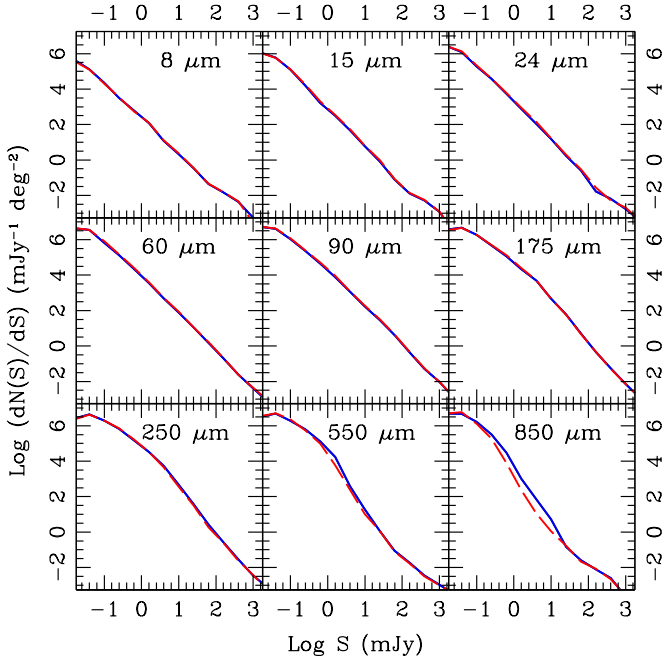}
    }
\caption{IR galaxy counts for \gal\ +\texttt{GRASIL}, comparing the full
computation (dashed red line) and the ANN approximation (continuous blue line).
The over-prediction in the ANN curve in the sub-mm is caused by a tiny
fraction high redshift models with some input neurons out of the range of
values covered by the trained net. See also Fig.\ \ref{fig:irlfhigh} and text
for explanation.} \label{fig:ircou}
\end{figure}

\begin{figure}
  \centerline{
   \includegraphics[width=0.5\textwidth]{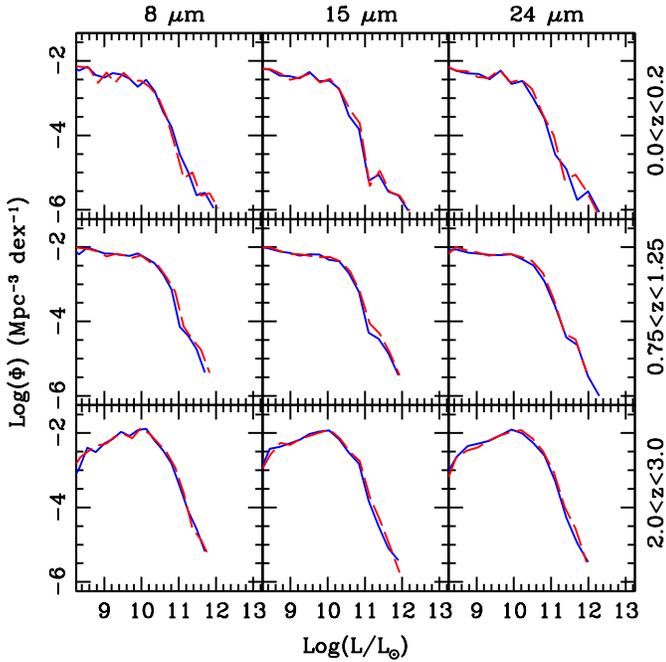}
    }
\caption{IR LF in the near and mid-IR bands and at three redshift bins for
\gal\ +\texttt{GRASIL}, comparing the full computation (dashed red line) and the ANN approximation (continuous blue
line).}
\label{fig:irlflow}
\end{figure}

\begin{figure}
  \centerline{
   \includegraphics[width=0.5\textwidth]{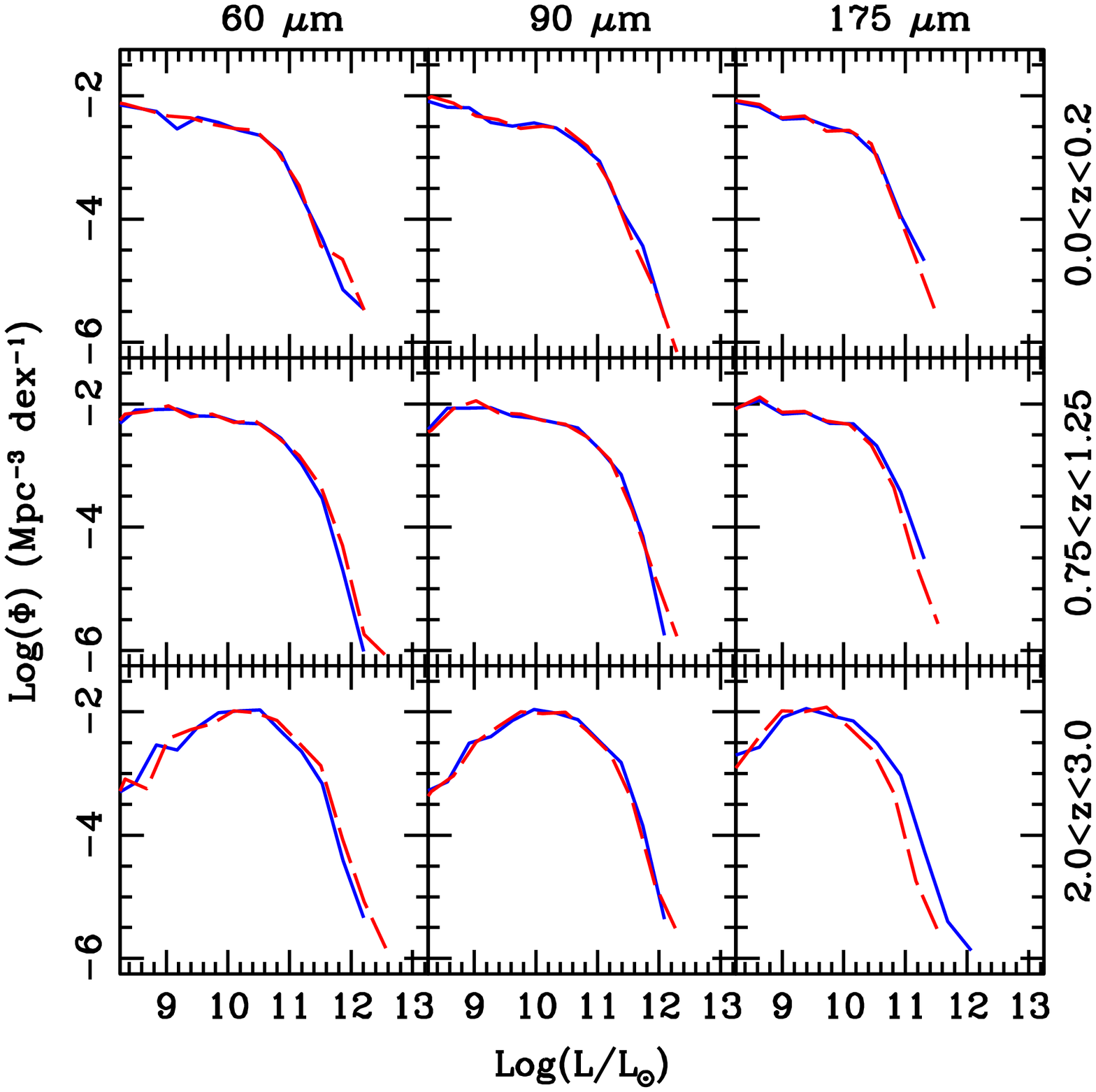}
    }
\caption{Same as Fig.\ \ref{fig:irlflow} but in the FIR. The mismatch in the ANN curve at high redshift is due to models
with some input neurons out of the range of values covered by the trained net. See text for explanation.}
\label{fig:irlfmid}
\end{figure}

\begin{figure}
  \centerline{
   \includegraphics[width=0.5\textwidth]{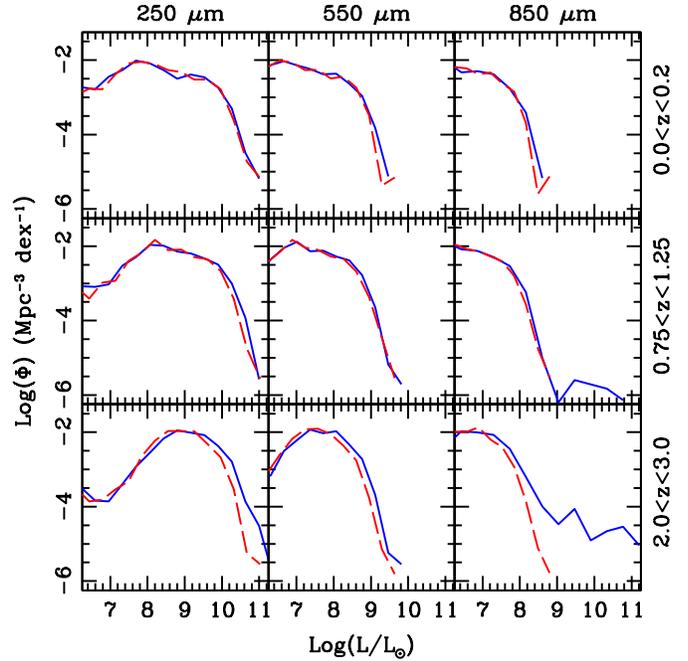}
    }
\caption{Same as Fig.\ \ref{fig:irlflow} but in the FIR and sub-mm. The mismatch in the ANN curve at high redshift is due
to models with some input neurons out of the range of values covered by the trained net. See text for explanation.}
\label{fig:irlfhigh}
\end{figure}

%%%%%%%%%%%%%%%%%%%%%%%%%

\section{Semi-analytic model: {\gal}}
\label{section:morgana}

To test the \gs-ANN, we make use of the
semi-analytic model {\gal}, developed by Monaco et al.\ (2007) and Fontanot et
al.\ (2007).  We refer the reader to these papers for details on the galaxy
formation model. Here we just recall its main features: the model
implements a sophisticated treatment of mass and energy flows between the
different gas phases (cold, hot and stars) and galactic components (bulge, disc
and halo), as well as a new treatment for gas cooling and infall (following
Viola et al.\ 2008). It also includes both a multi-phase description of star
formation and feedback (following Monaco 2004) and a self consistent
description of AGN activity and feedback (Fontanot et al.\ 2006).
In the following we use the {\gal} realization updated to the WMAP3 cosmology and Chabrier IMF as in Lo Faro et al.\ (2009).

Every model galaxy is represented by assuming a composite geometry including both
a spheroid and a disc component. Disc exponential profiles are computed
following the Mo et al.\ (1998) formalism: the spin parameter of the DM halo is
randomly extracted from a well defined distribution and the angular momentum is
conserved. Bulge sizes are computed assuming that the kinetic energy is
conserved in merger events (Cole 2000). The presence of a bulge is taken into
account when disc sizes are computed. {\gal} provides predictions for the star
formation history, metal enrichment, and mass assembly of each component
separately. This information is then interfaced with
\texttt{GRASIL} in order to predict the resulting SED from
the UV to the radio. The {\gal} realization we consider here is able to reproduce
the local and $z=1$ stellar mass function, the cosmic star formation history,
the evolution of the stellar mass density, the slope and normalization of the
Tully-Fisher relation for spiral discs, the redshift distribution and
luminosity function evolution for $K$-band selected samples, and, more
interestingly, the number counts of $850 \mu$m selected sources. Despite
these successes, the agreement between model predictions and observations of
the apparent ``downsized'' trend of galaxy formation is still under debate (see
Fontanot et al.\ 2007, 2009a for a complete discussion about ``downsizing''
trends and SAMs). It is well established that this model overpredicts the
number of faint, low-mass galaxies at $z<2$ (Fontana et al.\ 2006; Fontanot et
al. 2009a) and the space density of bright galaxies at $z<1$ (Monaco et al.\
2006). Moreover, it does not reproduce the observed levels of star formation
activity as a function of stellar and halo mass in the local universe (Kimm et
al.\ 2009). Despite being a particular choice among the various SAM
codes available in the literature, \gal\ is well suited for the
purposes of this paper, since it includes all the relevant physical
processes and it is representative of this technique (i.e.\ its
predictions are consistent with other codes, see e.g.\ Fontanot et al.\ 2009a).

%%%%%%%%%%%%%%%%%%%%%%%%%

\begin{figure*}
  \centerline{
   \includegraphics[width=\textwidth]{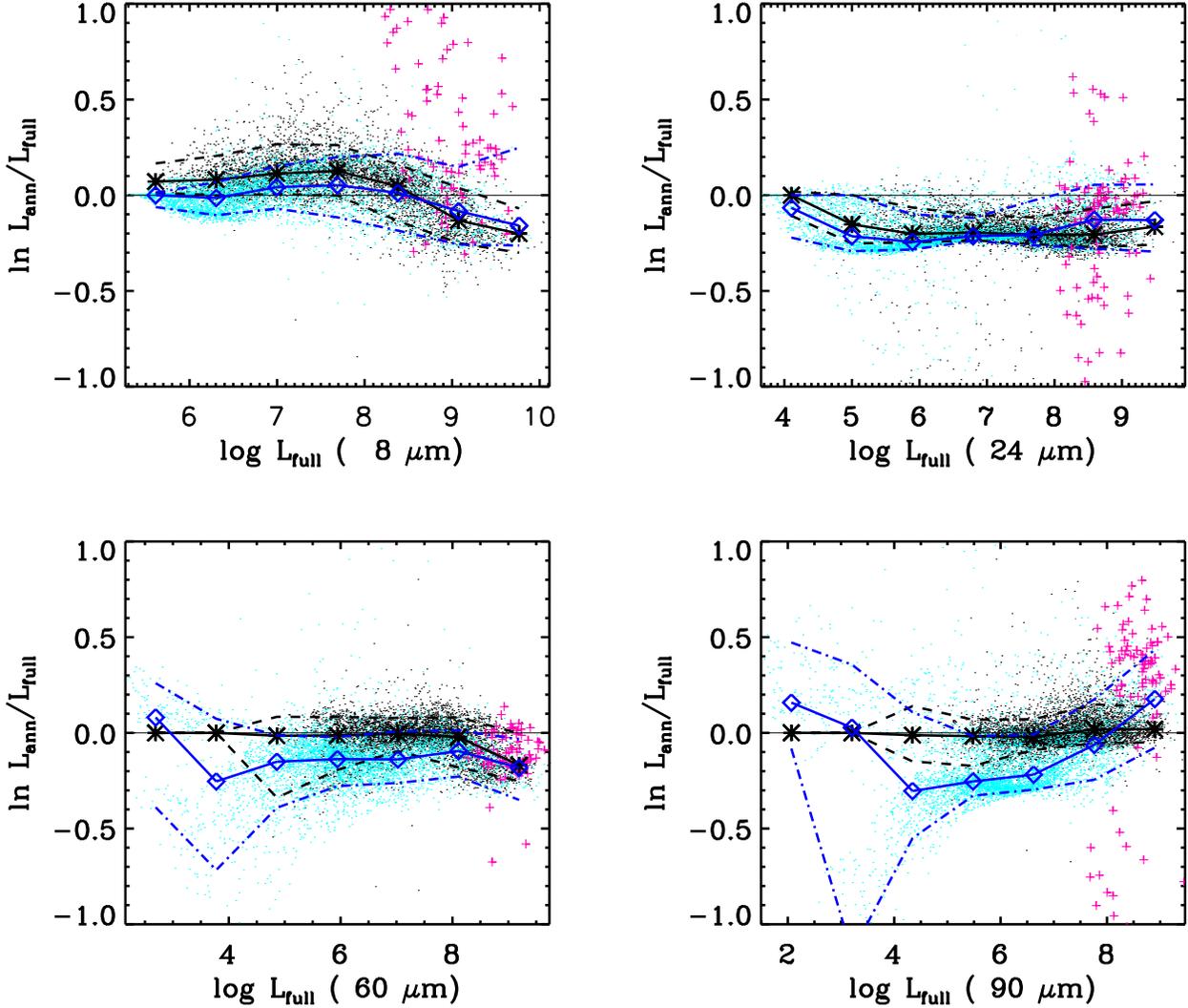}
    }
\caption{Logarithmic ratios of the ANN to the original monochromatic luminosities as a function of the latter (in $1e30$
erg/s/A), at 8, 24, 60, 90 $\mu$m for $10^4$ randomly extracted \gal\ models. The black continuous line with
asterisks and the short dashed lines are the median and the $0.1-0.9$ percentiles for those models (the black dots)
having all their input neurons within the range of the trained net. The blue continuous line with diamonds and the
dot-dashed lines are the same quantities for those models (the cyan dots) having at least one neuron out of the range of
the trained net. Among these, extreme models with a polar optical depth $\tau_p > 100$ at $1 \mu$m are highlighted by
pink crosses.}
\label{fig:rapl}
\end{figure*}

\begin{figure*}
  \centerline{
   \includegraphics[width=\textwidth]{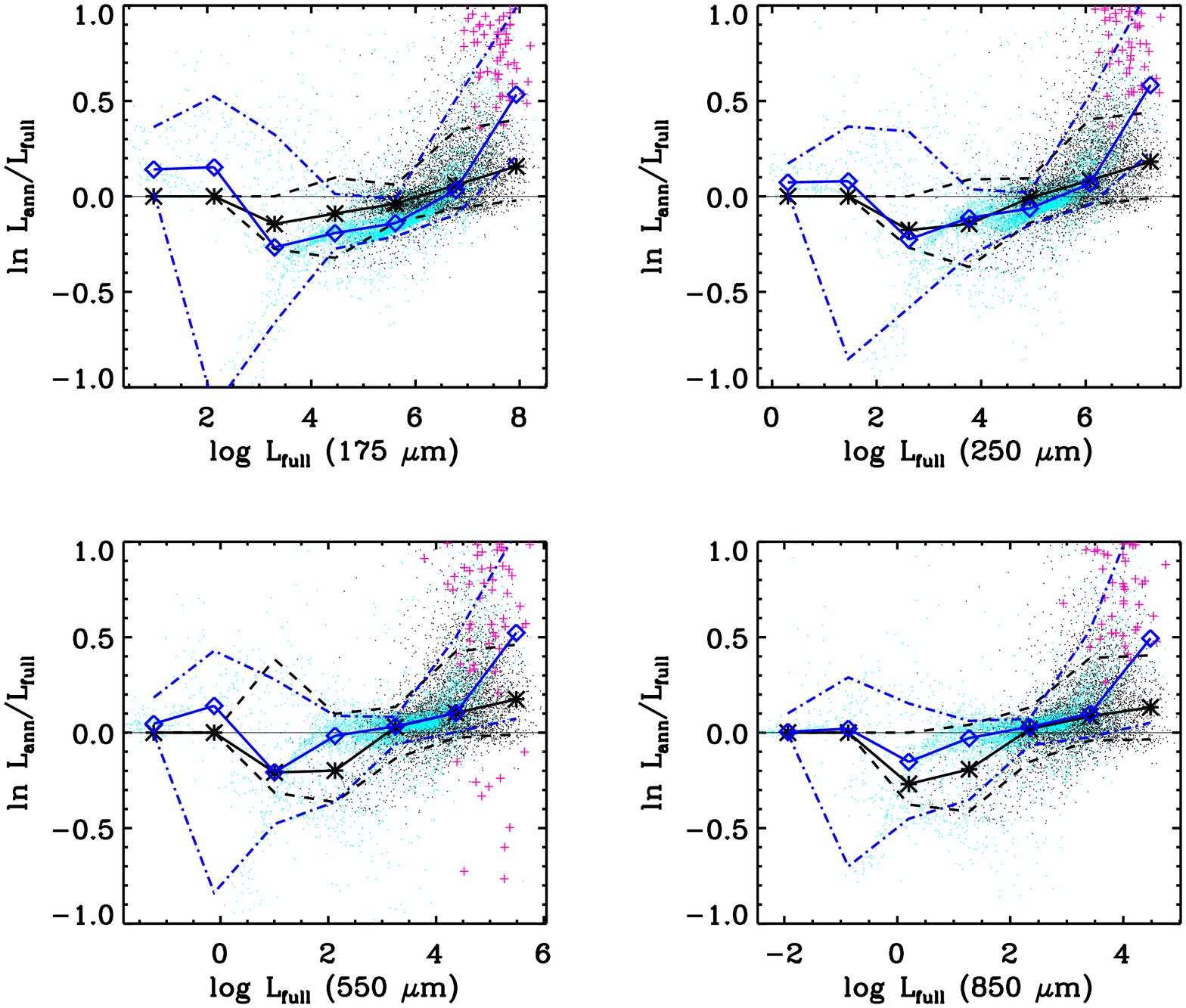}
    }
\caption{Same as Fig.\ \ref{fig:rapl}, but at 175, 250, 550, 850 $\mu$m.}
\label{fig:raph}
\end{figure*}

\section{Results}
\label{section:results}

As recalled in the Introduction, recent IR and sub-mm observations have allowed
to directly map the evolution of galaxies with cosmic time over large
cosmological volumes. In order to exploit this huge amount of multi-wavelength
data, we should be able to build large simulated volumes of galaxies and to
assign to each of them its SED. Another fundamental application is a
thorough exploration the parameter space of the SAM, practically unfeasible
with the full \texttt{GRASIL} computation.

We recall that the input for the trained net is computed with \texttt{CHE\_EVO}
and that we have blindly applied this net to {\gal}.
We run both the full and ANN versions of \texttt{GRASIL} on the same representative subsample of galaxies
extracted from a \gal\ realization of a cosmological box of 200 Mpc
size, following the same sampling procedure described in Fontanot et al. (2007).
Our final mock catalogue contains $95910$ galaxies in a redshift range
$0 \le z \le 8$.
We have then extracted the statistical quantities, counts and LFs,
from these two lightcones, and compared them (Figs. \ref{fig:ircou} to
\ref{fig:irlfhigh}).
This is done in several bands covering the whole near-IR to sub-mm range. Since
in this paper we are interested in only assessing the validity of the ANN method in
computing SEDs for SAMs, the comparison of this SAM with the data in IR
bands, and the ensuing tuning of parameters and/or physical ingredients, will
be investigated in another paper. In this work, we
keep the same parameter values as defined in Lo Faro et al.\ (2009).
Even if we found that $\sim 50\%$ of mock galaxies have at least one input neuron
out of the range of the trained net, only high-z starburst models with extreme
values of the optical depth caused prediction of the SED by the ANN so bad to
significantly affect sub-mm LFs and number counts. This is shown below.

Number counts are the most integrated observables used to test SAMs.
Therefore spectral differences between the two computing methods for individual
galaxies are smoothed out the most. In Fig.\ \ref{fig:ircou} we compare the
\texttt{MORGANA}+\texttt{GRASIL} galaxy counts for the full (dashed red line)
and the ANN (continuous blue line) computation in several bands: the {\it
Spitzer-IRAC} band at $8 \mu$m, the {\it ISO} 15, 90 and 175 $\mu$m bans, the
{\it Spitzer-MIPS} band at $24$ $\mu$m, the {\it IRAS} 60 $\mu$m, the {\it
Herschel-SPIRE} bands at $250$ and $550 \mu$m, and the {\it SCUBA} 850 $\mu$m
band. The two curves are almost superimposed at all wavelengths except in sub-mm bands, at
550 and even more at 850 $\mu$m.

In order to further check our method, in Figs.\ \ref{fig:irlflow},
\ref{fig:irlfmid} and \ref{fig:irlfhigh} we compare the ANN vs the full
computation for the redshift evolution of \gal\ luminosity functions, at the
same near- to far-IR and sub-mm bands as for the counts, in three redshift bins,
$z=0-0.2$, $0.75-1.25$, $2-3$. Although this statistic is less integrated, we
find again excellent agreement in most cases, but at high redshift and long
wavelengths, due to those same models selected by the sub-mm counts.

We have verified that the over-prediction in the ANN curve at $850 \mu$m is due to $\sim 60$ objects
(60 over 9678 at $z\sim 2.5$ , and 5 over 17230 at  $z\sim 1$)
characterized mainly by extreme values of the luminosity and of the dust optical depth:
for these models, in particular, the latter is far greater than the maximum values currently
available in the trained net used here. More specifically, all these galaxies
have a $1 \mu$m optical depth $\tau_p > 100$ (the maximum value of this neuron
in the trained net is 95), and actually most of them have $\tau_p$ between 200
and 300. Another characteristic for most of these models is an extremely high
normalized cirrus luminosity above the maximum currently available ($\log
(L_{Cir}/L_{\star,c})=0.1$). For these models the ANN provides an excess
FIR-summ cirrus SED for $\tau_p \sim 100$, and a catastrophic answer for
greater values. We explicitly note that these badly out-of-range models
do not affect other spectral ranges. Indeed, as in the observations,
sub-mm bands are particularly effective in selecting rare high-z,
very dusty and very luminous extreme mock galaxies, due to their favorable
negative k-corrections.

%%%%%%%%%%%%%%%%%%%

\begin{table}
\caption{RMS logarithmic error of the spectra predicted by the ANN with respect to the full computation at the
wavelengths of Figs.\ \ref{fig:rapl} and \ref{fig:raph}, for a sample of $10^4$ randomly extracted models. The errors are
given for the models with all neurons within the trained net, for those with at least one neuron out, and for the whole
sample.}
\label{tab:rmse}
\begin{center}
\begin{tabular}{cccc}
\hline
$\lambda$ [$\mu$m] & $e_{rms}$ (in) & $e_{rms}$ (out) & $e_{rms}$ (all) \\
\hline
8 & 0.157 & 0.175 & 0.167 \\
24 & 0.229 & 0.255 & 0.243 \\
60 & 0.107 & 0.268 & 0.205 \\
90 & 0.104 & 0.311 & 0.233 \\
175 & 0.202 & 0.347 & 0.285 \\
250 & 0.235 & 0.350 & 0.298 \\
550 & 0.257 & 0.310 & 0.285 \\
850 & 0.240 & 0.533 & 0.414  \\
\hline
\end{tabular}
\end{center}
\end{table}
%%%%%%%%%%%%%%%%%

To investigate this point, we have randomly extracted from the \gal\ catalogue
a sample of $10^4$ models, over the $0$ to $8$ redshift range and compared
directly the spectral behavior of the ANN vs the full computation. In Figs.\
\ref{fig:rapl} and \ref{fig:raph} we show the logarithmic ratios of the ANN to
original \gal\ + \texttt{GRASIL} monochromatic luminosities versus the latter,
at different wavelengths from 8 to 850 $\mu$m. The dots are the models (black
for models with all the neurons within the range of the trained net, cyan for
those models with at least one neuron out range). The line with asterisks and
the short-dashed lines are respectively the median and the $0.1-0.9$
percentiles for in-range models. The same quantities for out-of-range models
are the line with diamonds and the dot-dashed lines. The statistical noise
appearing in some plots at low luminosities are due to having a few points
falling within the luminosity bins used to evaluate the median and the
percentiles. The rms logarithmic errors at the different wavelengths are in
Table \ref{tab:rmse} ($e_{rms}=\sqrt(\sum^n(\ln L_{ann}/L_{full})^2/n)$). There is a
general dependency of the dispersion on the luminosity. In particular, the
dispersion of the out-of-range models at the highest luminosity bins is
dominated by extremely dusty models. In fact we have highlighted with crosses
the models with a polar $1 \mu$m optical depth $\tau_p
> 100$. As we found previously for the sub-mm counts and LFs, also among this
random sample, the models having a catastrophic answer by the ANN are all
characterized by extreme values of $\tau_p$. In order to safely use (far-)IR
data to constrain the SAM, it will be therefore advisable to enlarge the range
of values covered by the training so to encompass also those extreme models
predicted by the SAM but difficult to expect a priori.

%%%%%%%%%%%%%%%%%%%%%%%%%%%

\section{Summary and conclusions}
\label{section:concl}

It is well known that the numerical solution of the radiative transfer is time expensive, and
this is particularly a concern for semi-analytical galaxy formation models,
whose aim is to simulate large cosmological volumes to compare with
multi-wavelength surveys, or to thoroughly explore and constrain the model
parameter space. On the other hand, the typical way to circumvent the computing
time problem has been to rely on template SEDs. This can be misleading,
since the templates, typically observationally based, can become very different
from those that mock galaxies would have given their features, as predicted by
the SAM (geometry and SF history).

We have implemented in \gs\ an artificial neural network algorithm in order to
significantly speed up the computation of the IR SED, which is bottle-neck of
the full SED computation. The ANN is able to excellently approximate the full
computation, but with a gain in CPU time by $\sim 2$ orders of magnitude.

The main points of our work are as follows:

\begin{itemize}

\item In Paper I we have proven that the ANN algorithm was suited in the
    cases of the simplified pure bulge or pure disk geometries. Here we
    have further extended the ANN to the mixed bluge+disk geometry, suited
    for fully fledged SAMs. This required to account for additional input
    neurons to fully define the IR SEDs.

\item An important characteristics of the ANN we have implemented is that
    it is of general use, i.e. it is not linked to a particular SAM.
    Indeed, the input neurons are defined as those quantities effectively
    determining the shape of the SEDs from the star forming clouds and the
    diffuse medium. Therefore, the training of the ANN can be performed
    with any model and then applied to other models. It is only advisable
    to cover the range of values of the input neurons encompassed by the
    model the net is applied to.

\item We have checked the algorithm with the SAM \gal. We made a blind
    test, by applying a net trained with \texttt{CHE\_EVO} (a standard
    chemical evolution model) covering a large range of input values, to a
    \gal\ catalogue. We ran the full and the ANN \gs\ on this catalogue,
    and compared the results for the galaxy counts and evolution of the LFs
    in several near-IR to sub-mm bands. The agreement is excellent except
    in the sub-mm at high redshift. We have checked that the disagreement at long wavelength
    is caused by a tiny fraction of models with some sensible input
    neurons out of the range of the trained net. These are extreme
    starbursting models with very high optical depths, favorably
    selected by sub-mm observations.

\item Since counts, and to a lesser extent LFs, are integrated quantities
    for which small spectral differences are smoothed out, we have also
    performed a check on the spectral differences between the full and the
    ANN, by randomly selecting $10^4$ models from the \gal\ catalogue. We
    compare the median and the percentiles of the spectral luminosity
    ratios $L_{ANN}/L_{full}$, separately for those models having all the
    neurons within the range of the trained net, and those with at least
    one neuron out. We confirm the previous findings for the goodness of
    the algorithm, and also that the large spread of out-of-range models
    take place at high luminosities and is ascribed to extremely dusty
    galaxies produced by \gal\ at high-z.

\end{itemize}

We conclude that the performance of the ANN algorithm is perfectly suited for
the purpose of simulating galaxy lightcones with SAM and exploiting data from
multi-wavelength cosmological surveys, and for effective exploration of SAMs parameter space.
The fact that this implementation is
independent of the specific model used for generating the training set, will
allow a single large training to be applied to different SAMs.
The results
are totally safe, at least for statistical purposes such as LFs and counts,
provided that the training covers reasonably well the range of input neurons in
the application. This condition is easy to check. In the event it were not
satisfied, possible solutions are to broaden the training, or to run the full
\gs\ for a few problematic mock galaxies. The best choice depends on the
frequency of the latter and on the desired application. In this work we
have used \gal\ as a realistic test for the ANN method. The comparison of \gal\
with IR data, and the likely tuning of its parameters and/or physical
ingredients, exploiting this technique, will be the subject of a future work.

%%%%%%%%%%%%%%%%%%%%%

\section*{Acknowledgments}

FF acknowledges financial support from the Klaus Tschira Foundation.
Some of the calculations were carried out on the "Magny" cluster of the
Heidelberger Institute f\"ur Theoretische Studien.

{}

\clearpage

\end{document}